\documentclass[english,aps,floatfix,superscriptaddress,showpacs,amsfonts,amssymb,superscriptaddress]{revtex4}
\usepackage{color}

\usepackage{graphicx}
\usepackage{amsmath}
\usepackage{latexsym}
\usepackage{epstopdf}
\usepackage{amsmath}
\usepackage{amssymb,amsmath}
\usepackage{multirow}
\usepackage{mathtools}
\newcommand{\be}{\begin{equation}} \newcommand{\ee}{\end{equation}}
\newcommand{\ba}{\begin{eqnarray}} \newcommand{\ea}{\end{eqnarray}}
\newcommand{\bear}{\begin{eqnarray*}} \newcommand{\eear}{\end{eqnarray*}}
 
\newcommand{\rf}[1]{(\ref{#1})}

\begin{document}


\title {
 Cyclic representations  of the    periodic Temperley Lieb
  algebra, \\
 complex
Virasoro representations   and  stochastic processes}
\author{Francisco C. Alcaraz}
\email{alcaraz@ifsc.usp.br}

\affiliation{ Instituto de F\'{\i}sica de S\~{a}o Carlos, Universidade de S\~{a}o Paulo, Caixa Postal 369, 13560-970, S\~{a}o Carlos, SP, Brazil}

\author{Arun Ram} 
\email{ aram@unimelb.edu}
\affiliation{Department of Mathematics and Statistics, 
 University of Melbourne, Parkville, VIC 3010, Australia} 

\author{Vladimir Rittenberg}
\email{ vladimir@th.physik.uni-bonn.de}

\affiliation{Physikalisches Institut, Universit\"at Bonn,
  Nussallee 12, 53115 Bonn, Germany}

\date{\today{}}
\date{\today}


\begin{abstract}
An $N$${L} \choose {L/2}$ dimensional representation of the periodic Temperley-Lieb
algebra $TL_L(x)$ is  presented. It is also a representation of the
cyclic group $Z_N$. We choose $x = 1$ and  define a Hamiltonian 
 as a sum of
the generators of the algebra acting in this representation. This
Hamiltonian gives the time evolution operator of a stochastic process. In the
finite-size scaling limit, the spectrum of the Hamiltonian contains
 representations of the Virasoro algebra with complex highest weights. The $N = 3$ case is
discussed in detail.
One discusses shortly the consequences of the existence of complex
Virasoro representations on the physical properties of the systems.
\end{abstract}

\maketitle

The periodic Temperley-Lieb algebra $PTL_L(x)$ was introduced by Levy \cite{LEV}
in 1991 in order to explain some regularities observed in the  spin
1/2 XXZ quantum chain with periodic boundary conditions \cite{AGR}. The algebra has $L$
generators and depends on a parameter $x$. Various quotients of this algebra
were studied by Martin and Saleur \cite{MAS}. A renewed interest in the 
$PTL_L(x)$ 
appeared in the last few years in the context of logarithmic conformal field
theory \cite{MPR} and \cite{GSR}. 
Lately stochastic processes describing nonlocal
asymmetric exclusion processes (NASEP) were studied using representations of
the same algebra \cite{ALR}.

  In the present Letter we consider a new quotient and cyclic representations
of the algebra. As usual one can define a Hamiltonian expressed in terms of
generators of the $PTL_L(x)$. If one takes $|x| < 2$, 
use cyclic representations,
and consider the finite-size limit of the spectra of the Hamiltonian, we
show that they can be expressed in terms of complex representations of the
Virasoro algebra. To our knowledge, it is for the first time that such
representations are seen in physical problems.

  The $PTL_L(x)$ algebra has $L$ generators $e_k$  ($k = 1,2,\ldots,L$) 
satisfying the
relations \cite{LEV}:
\be \label{e1}
  e_k^2 = xe_k, \quad  e_ke_{k\pm 1}e_k = e_k, \quad [e_k,e_l] = 0 \quad (|k-l| > 1), \ee
and $e_{k+L} = e_k$.

  For simplicity we take $L$ even. We consider the quotient:
\be \label{e2}
  (AB)^N A = A,
\ee
where
\be \label{e3}
 A= \prod_{j=1}^{L/2} e _{2j}, \quad  B = \prod_{j=0}^{L/2-1}  e_{1 + 2j}.
\ee

In the definition \rf{e2} $A$ and $B$ can be interchanged. 
The case $N = 1$ is one
of the quotients of reference \cite{MAS} 
\be \label{ez}
ABA=\alpha A 
\ee
with $\alpha=1$. Representations of the quotient \rf{ez} in terms of 
quantum chains were discussed in \cite{LEV} and in  \cite{MSA}. Notice that 
choosing $\alpha=\exp(i 2\pi r/N)$ with $r=0,1,2,\ldots,N-1$ in \rf{ez} one 
obtains $N$ independent representations of the quotient \rf{ez}. In what 
follows we present different representations of the same quotient.

\begin{figure}
\centering
\includegraphics[angle=0,width=0.3\textwidth] {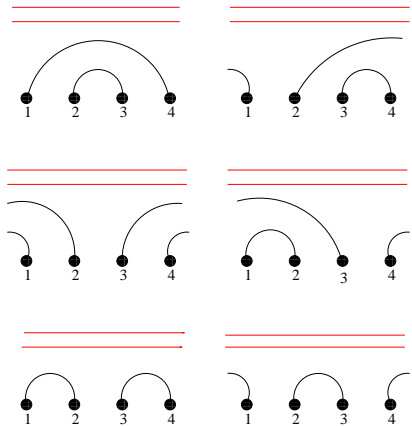}
\caption{ 
 The six link patterns configurations for $L = 4$ sites on a cylinder
and two circles without sites (noncontractible loops). The open arches and
circles meet behind the cylinder.}
\label{fig1}
\end{figure}

  We now show that the $PTL_L(x)$ has $Z_N$ cyclic link representations ($Z_N$ is 
the cyclic group of order $N$).   
Consider $N$
copies ($n = 0, 1, 2,\ldots,N - 1$) of periodic link patterns. 
Each link pattern
is one of the 
${L} \choose  {L/2}$ 
 configurations of nonintersecting arches 
joining $L$ sites
on a circle. One can think of having the circle on a cylinder. Each copy $n$ 
is
labeled by $n$ circles on the same cylinder with no sites on them
(noncontractible loops). In Fig~\ref{fig1} we show the 6 configurations for $L = 4$
and $n = 2$.  The open arches and the
circles join in the unseen side of the cylinder.
  
With a few exceptions, the generators $e_k$ act on the configurations of a
given copy  in the standard way \cite{MAR}.
\begin{figure}
\centering
\includegraphics[angle=0,width=0.3\textwidth] {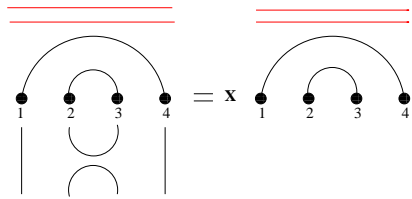}
\caption{ 
 The action of the $e_2$ generator acting on the bond between the sites
2 and 3 which are not the end-points of an arch of the size of the system.
$L = 4$, $n = 2$ in the figure.}
\label{fig2}
\end{figure}

\begin{figure}
\centering
\includegraphics[angle=0,width=0.3\textwidth] {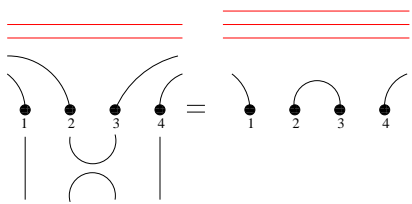}
\caption{ 
The action of the $e_2$ generator acting the bond between the sites 2
and 3 which are the end of an arch of the size of the system $L = 4$. A new
circle is created on the cylinder and one moves from the copy $n = 2$ to the
copy $n = 3$.}
\label{fig3}
\end{figure}

In Fig.~\ref{fig2} we show the action of $e_2$ on one of the configurations 
shown in
Fig.~\ref{fig1}. The factor $x$ appears due to a contractible loop. 
The exceptions
occur if on the copy $n$ one considers a configuration having an arch of the
size $L$ of the system and if the generators acts on the bond between the
two ends of the arch (see Fig.~\ref{fig3}). The action of $e_2$ on the third 
configuration
in Fig.~\ref{fig1} produces a new circle and therefore gives a 
configuration in the copy
$n = 3$.

  What we have seen in this example is a general phenomenon. If a
generator acts on a bond connecting two sites which are the end-points of
an arch of length $L$ of the copy $n$, one obtains a configuration belonging to
the copy $n + 1$.

\begin{figure}
\centering
\includegraphics[angle=0,width=0.3\textwidth] {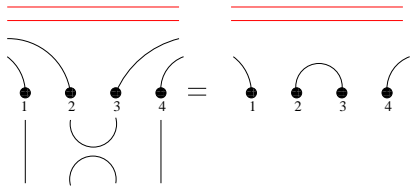}
\caption{ 
One takes $N = 3$. The action of the generator $e_2$ described in 
Fig.~\ref{fig3} 
is changed depending of the quotient one chooses. In the figure one shows
the choice of the quotient of Eq.~\rf{e4}. One doesn't change the copy which
stays $n = 2$.}
\label{fig4}
\end{figure}

  In order to get a finite-dimensional representation of the algebra, one
has to take a decision. The simplest one is to identify the copy $N$ 
with the copy $N-1$. This possibility is illustrated in Fig.~\ref{fig4} for the case
$N = 3$ and $L = 4$.

It is easy to check that one obtains in this way not a representation of
the quotient \rf{e2} but of a different quotient:
\be \label{e4}
(AB)^{N} A = (AB)^{N-1} A
\ee
  
Representations of the quotient \rf{e4} might be interesting in their own
right but we didn't study them here.

\begin{figure}
\centering
\includegraphics[angle=0,width=0.3\textwidth] {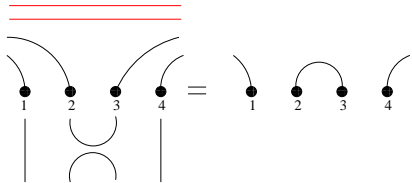}
\caption{ 
 Same as in Fig.~\ref{fig4} choosing the quotient given by Eq.~\rf{e2}. 
From the
copy $n = 2$ one moves to the copy $n = 0$ in order to get a representation
with the symmetry $Z_3$.}
\label{fig5}
\end{figure}

  In order to obtain representations of the quotient \rf{e2} we have to
identify the copy $N$ not with the copy $N - 1$ but with the copy $n=0$ (no
noncontractible loops). See Fig.~\ref{fig5}  for $N = 3$ and $L = 4$. 
By adding circles without sites this representation is also a  representation
of the cyclic group  $Z_N$. One can show \cite{GL} that this 
representation is reducible. It splits into $N$ representations defined by 
the quotients \rf{ez} with $\alpha = \exp(i2\pi r/N)$. 

In what follows we consider the application of cyclic representations to
stochastic processes \cite{ALR} taking $x = 1$. The Hamiltonian
\be \label{e5}
  H = \sum_{k=1}^L(1 - e_k), 
\ee
gives the time evolution of the probability distribution function defined in
the configuration space of the $N$ copies of link patterns each containing
$ {L}\choose {L/2}$ configurations. A detailed discussion of the spectra of $H$ will be
presented elsewhere \cite{APR}. For the remaining of this Letter we consider only even values of $L$.

  We first recall the known case $N = 1$. We use the spin representation of
the $PTL_L(1)$ \cite{LEV,MSA}:
\ba \label{e6}
e_k &=&\sigma_k^+\sigma_{k+1}^- +
\sigma_k^-\sigma_{k+1}^+ + \frac{1}{4} (1-\sigma_k^z\sigma_{k+1}^z) +i\frac{\sqrt{3}}{4} (\sigma_{k+1}^z -\sigma_{k}^z), \quad k=1,2,\ldots,L-1,
\nonumber \\
e_L &=&e^{i\frac{2\pi}{3}}\sigma_L^+\sigma_{1}^- +
e^{-i\frac{2\pi}{3}}\sigma_L^-\sigma_{1}^+ + \frac{1}{4} (1-\sigma_L^z\sigma_{1}^z) +i\frac{\sqrt{3}}{4} (\sigma_{1}^z -\sigma_{L}^z).
\ea

In the scaling limit, the scaling dimensions $\{x\}$ are obtained from the leading behavior of the energy-gap amplitudes $E = 2\pi v_s x/L$, where $v_s=3\sqrt{3}/2$ is the sound velocity.
The spectrum of $H$ in the link representation is contained in the 
$S^z =\sum_{k =1}^L\sigma^z_k = 0$ sector and is known.  The scaling dimensions associated to the eigenstates with 
momenta $P=2\pi p/L$ (mod. $\pi$) ($p=0,\pm1,\pm2,\ldots$), are \cite{AR1,SAL}
\be \label{e7}
x = 3/4(1/3 + s)^2 - 1/12 + m + m', \quad 
p = m - m'.
\ee
where $s,m,m'=0,\pm1,\pm2,\ldots$.

The lowest excitation is obtained if one takes $ s = -1, m=m'=0$
\be \label{e8}
x ^0_0(1) = 1/4 = 0.25, \quad p=0.
\ee
The explanation of the notation $x^0_0(1)$ will be given in few lines.

  If $N \neq 1$, the states are separated not only by the momenta but also by
the $Z_N$ representation $\exp(i2\pi r/N)$  to which they belong ($r = 0, 1, 2,\ldots,N-1$). 
The $r = 0$
states for example, are obtained by taking the sum of the same link
configuration in all the $N$ copies. We will denote by $x^r_p(i)$ 
($i=1,2,\ldots $) the scaling
dimensions associated to the $i$-th lowest energy in the sector of momentum $P=2\pi p/L$ (mod $\pi$) and $r$ representation of $Z_N$. 
In
what follows we present some results for the case $N = 3$.

Although it is known \cite{DPR} that the system is integrable but the 
calculations are tedious, we have studied the finite-size scaling spectra numerically using up to $L =
30$ sites. 
We separate the vector space into disjoint sectors labelled by the momentum $P$ and the index $r$ of the representation $Z_N$. The lowest energy in each 
sector is calculated by the power method. 
The ground state of $H$ which corresponds to the stationary
state of the stochastic process, corresponds to the eigenvalue zero.
The eigenfunction is in the $p = 0, r = 0$ sector and shows no new
combinatorial properties beyond those known from the $N =1$  case \cite{ZZZ}. 
One
relevant result is that the entire spectra related to the scaling dimensions $\{x^0_p(i)\}$ coincide with the known spectra of
the $N = 1$ representation. In order to show the precision of our procedure, we
have estimated the scaling dimension \rf{e8} just from the energy gap 
for a $L = 30$
lattice (no extrapolations using different sizes!) and got 0.24976220.

  Taking $N = 3$ we looked at the spectra in the $r = 1$, and $p = 0$  
sectors and got a
surprise. The extrapolants\cite{VBS} for the two first excited levels gave the
following complex values:
\be \label{e9}
x^1_0(1) = 0.03905  + 0.08753\;i, \quad  x_0^1(2)=0.14908 -0.11806\;i.   
\ee

  In order to check if these results have anything to do with conformal
invariant spectra, we looked at the $r = 1, P = 2\pi/L$ (mod. $\pi$) spectrum. If the
finite-size scaling limit of the spectra  are given by Virasoro representations with  
a complex highest weight, one should expect $x^1_1(i) = x^1_0(i) + 1$ ($i=1,2$). This is
indeed the case since we get:
\be \label{e10}
x^1_1(1) = 1.0391  + 0.08755\;i, \quad     x_1^1(2)=1.149 - 0.11806\;i.   
\ee

In order to illustrate the precision of the estimates of the scaling
dimensions, in table 1 and 2  we give their measured values for different lattice
sizes. One can see that the data converge very nicely. In the $r = 2$
sector one obtains the complex conjugate values of \rf{e9} and \rf{e10}: 
$x^2_p
= (x^1_p)^{\dag}$.
The very existence of Virasoro representations 
is a remarkable fact since the transitions from one copy to another is a 
highly nonlocal operation.

\begin{table}[htp]
\caption{ Numerical estimates for  the lowest two scaling dimensions 
appearing in the
sector $r=1$ and momentum $0$ (mod $\pi$). In the last line of the table we show the
results obtained by the van den Broeck and Swartz extrapolants (VBS). The complex conjugated dimensions  appear in the sector $r=2$}
\begin{center}
\label{tab1}
\begin{tabular}{lcc}

 $L$ & $x_0^1(1)$ &$x_0^1(2)$ \\  
\hline
   6 & $0.0411337612 +  0.0893222227\;i$ &$0.1522015017 -0.1522015017\;i$ \\
  10 & $0.0398186156 +  0.0883428114\;i$ &$0.1501651591 -0.1171087213\;i$ \\
  14 & $0.0394510154 +  0.0880912156\;i$ &$0.1496233621-0.1175177089\;i$ \\
  18 & $0.0392969971 +  0.0879921416\;i$ &$ 0.1494051829-0.1177070065\;i$ \\
  22 & $0.0392178087 +  0.0879436454\;i$ &$0.1492965628 -0.1178110010\;i$ \\
  26 & $0.0391716570 +  0.0879165117\;i$ &$0.1492349425 -0.1178746350\;i$ \\
  30 & $0.0391423702 +  0.0878998853\;i$ &$0.1491967343 -0.1179165804\;i$ \\
  $\infty$ & $0.039050 +  0.087853\;i$ & $0.149085  -0.11806\;i$

\end{tabular}
\end{center}
\end{table}

\begin{table}
\caption{ Numerical values of the lowest two scaling dimensions appearing in the
sector $r=1$ and momentum $2\pi/L$ (mod $\pi$)  (the complex conjugated 
dimensions  appear in the sectors with  $r=2$ and momentum $2\pi/L$ (mod $\pi$). In the last line of the table we show the results obtained by the VBS  extrapolants.}

\begin{center}
\label{tab2}
\begin{tabular}{lcc}

 $L$ & $x_1^1(1)$  & $x_1^1(2)$ \\ 
\hline
   6 & $0.8954988326+0.0427352699\;i$ & $0.9485861617 -0.0543306613\;i$ \\  
  10 & $0.9856155271+0.0697674849\;i$ & $1.0727332613 -0.0917528656\;i$ \\  
  14 & $1.0114751102+ 0.0781595719\;i$ & $1.1092111061 -0.1038306091\;i$ \\
  18 & $1.0222671807 + 0.0818051754\;i$ & $1.1246236272 -0.1091468356\;i$ \\
  22 & $1.0277701176 + 0.0837133906\;i$ & $1.1325471271-0.1119468246 \;i$ \\
  26 & $1.0309499010 +0.0848374121\;i$ & $1.1371534904 -0.1136018590 \;i$ \\
  30 & $1.0329516135 +0.0855557589\;i$ &$1.1400672494 -0.1146618390 \;i$ \\ 
  $\infty$ & $1.0391  +0.0878\;i$  & $1.149 - 0.11806 \;i$

\end{tabular}
\end{center}
\end{table}

  Notice that the scaling dimensions \rf{e9} have a smaller real part than
the value \rf{e8}. This observation has physical consequences. If we consider
a local observable, using the mappings of the link patterns into Dyck
paths, charge particles or particles-vacancies configurations \cite{ALR},
for large systems, the approach to the stationary state will be oscillatory.
As far as we know, it is the first time that such a phenomenon can be
observed since normally the imaginary part of the energy levels decreases
faster with $L$ than the real part. There are obviously consequences for
the correlation functions too. We should stress that the stochastic process with the evolution operator \rf{e5} takes place in the $N$${L}\choose{L/2}$ 
dimensional vector space which is a representation of $Z_N$ and not in the 
independent copies \rf{ez} with $\alpha=\exp(i\pi r/N)$. The spectra are 
related but one has to have in mind that in a stochastic model the 
wavefunctions must have real nonnegative coefficients and that the various 
sectors are mixed.
 
\begin{figure}
\centering
\includegraphics[angle=0,width=0.40\textwidth] {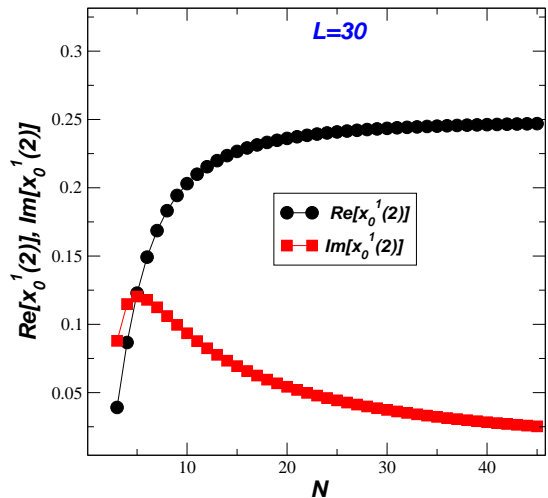}
\caption{ Real part (black) and imaginary part (red) of the estimated value of the scaling dimension $x_0^1(2)$, as a function of $N$  for the lattice size 
$L=30$.}
\label{fig6}
\end{figure}

We would like to mention that we have also looked at the variation of the 
lowest excited states
 with $N$ keeping $r = 1$. The data for the second level with momentum zero (mod $\pi$) are shown in Fig.~\ref{fig6}. 
One sees that
increasing $N$, the real part approaches the value \rf{e8} and that the 
imaginary part gets
smaller and smaller. This is not to say that this  scaling 
dimension can't be
found for another value of $r$ but the consequence of the data shown in 
Fig.~\ref{fig6} is that in
 the large $N$ limit, the scaling dimension 1/4 will be found at least 3 
times ($r = 0,
1$, and 2).  
We have not looked at the possible existence of Jordan cells in the
spectrum \cite{MDA}.

In \cite{APR} we will give the partition function for each sector $r$ and 
for any parameter $x$ of the definition of the algebra \cite{LEV}.
The case $x = 0$ is especially interesting 
since in this case, the
Hamiltonian is related to the transfer matrix of a classical system 
of $N$ colored
interacting polymers on a cylinder generalizing the known case $N = 1$ 
\cite{MPR}.

\section{Acknowledgements} 

We would like to thank Alexi Morin-Duchesne, Paul Pearce, Pavel Pyatov, 
Hubert Saleur  and Paul Martin for 
discussions. VR is grateful to Jan de Gier for the invitation 
at the Melbourne University where part of this research was done. This 
work was supported in part  
by the Australian Research Council (Australia),  FAPESP and CNPq (Brazilian Agencies).

\end{document}